\documentclass[12pt,fleqn]{article}
\usepackage{amsfonts}
\usepackage{amsmath}
\usepackage[a4paper,margin=1in]{geometry}
\usepackage[nodisplayskipstretch]{setspace}
\usepackage{graphicx}
\usepackage{float}
\usepackage{color}
\usepackage{subcaption}
\usepackage{multirow}
\usepackage{pgffor}
\usepackage{mathtools}
\usepackage{amssymb}
\usepackage{eurosym}
\usepackage{natbib}
\usepackage{hyperref}
\usepackage{mathtools}
\usepackage{amsfonts}
\usepackage{eurosym}
\usepackage{amssymb}
\usepackage{amsmath}
\usepackage{graphicx}
\usepackage{float}
\usepackage{color}
\usepackage{natbib}
\usepackage{hyperref}
\usepackage[a4paper, margin=1in]{geometry}
\usepackage[nodisplayskipstretch]{setspace}
\usepackage{multirow}
\usepackage{subcaption}
\usepackage{pgffor}
\usepackage{booktabs}
\usepackage{adjustbox}
\usepackage{graphicx}

\hypersetup{
colorlinks=true,
linkcolor=black,
citecolor=black,
filecolor=magenta,
urlcolor=black}
\newtheorem{theorem}{Theorem}
\newtheorem{example}{Example}

\newtheorem{assumption}{Assumption}

\newtheorem{proposition}{Proposition}
\newtheorem{corollary}{Corollary}

\numberwithin{lemma}{section}

\newcommand{\Rom}[1]{\uppercase\expandafter{\romannumeral #1\relax}}
\newcommand{\rom}[1]{\lowercase\expandafter{\romannumeral #1\relax}}


\graphicspath{{figs/}}

\begin{document}

\title{Properties of the Conditional Likelihood Ratio Test under Discrete Approximation \thanks{
We thank Marinho Bertanha, Marcelo Fernandes, Raul Guarini, Michael Jansson and Lucas Lima for helpful comments, Isaiah Andrews for productive discussions and for sharing numerical codes, and Guilherme Exel and Marcos Willi for excellent research assistance. This study was financed
in part by the Coordena\c{c}\~{a}o de Aperfei\c{c}oamento de Pessoal de N%
\'{\i}vel Superior - Brasil (CAPES) - Finance Code 001. This research was also supported in part by the University of Pittsburgh Center for Research Computing and Data, RRID:SCR\_022735, through the resources provided. Specifically, this work used the H2P cluster, which is supported by NSF award number OAC-2117681. We also gratefully
acknowledge the research support of CNPq and FAPERJ.}}
\date{\today}
\author{ Marcelo J. Moreira \thanks{
{\scriptsize Postal address: Praia de Botafogo, 190, Rio de Janeiro, RJ,
22250-145, Brazil. Email: \href{mailto: moreira.marcelo.j@gmail.com}{%
moreira.marcelo.j@gmail.com}. }} \\
{\footnotesize FGV/EPGE} \and Mahrad Sharifvaghefi \thanks{{\scriptsize %
Postal address: 230 S Bouquet St., Pittsburgh, PA, 15260, USA. Email: \href{mailto:sharifvaghefi@pitt.edu}%
{sharifvaghefi@pitt.edu}. }} \\
{\footnotesize University of Pittsburgh} }
\maketitle

\begin{abstract}
The conditional likelihood ratio (CLR) test is a valuable tool for inference under weak identification,
with appealing theoretical properties in both linear and nonlinear settings.
Its implementation nevertheless requires minimizing a non-convex objective function, a difficulty long recognized even in the linear IV setting.
While grid-based methods that provide a practical approximation may perform well in particular designs, such procedures do not guarantee that the resulting test preserves the theoretical properties of the CLR test uniformly across a class of data-generating processes. This paper examines the
implementation challenges and their consequences for test size and power. In
the linear IV settings, we contrast the grid-based  method with the polynomial approach of \Citet*{%
MoreiraNeweySharifvaghefi24}, which guarantees global minimization and aligns computation with
the theoretical properties of the CLR test.
\end{abstract}

{\setstretch{1.1} }

\pagebreak

\section{Introduction}
In linear instrumental variable (IV) models with homoskedastic errors, \Citet*{Moreira03} develops the conditional likelihood ratio (CLR) test, which uses a sufficient statistic for the nuisance parameter to construct a valid critical value function, even in the presence of weak identification. The CLR test
has notable theoretical properties. For cases with a single endogenous
variable and homoskedastic errors, \citet*{AndrewsMoreiraStock06}
demonstrate that the CLR test is nearly uniformly most powerful within an
important class of tests. Furthermore, it is not subject to the low-power behavior that \citet{MoreiraRidderSharifvaghefi26} identify for other tests that are robust to heteroskedastic and autocorrelated (HAC) errors in the IV literature, including the Lagrange multiplier (LM) and conditional quasi-likelihood ratio (CQLR) tests.

\citet*{MoreiraMoreira19} provide the CLR statistic in linear IV models with HAC
errors. \citet*{MoreiraRidderSharifvaghefi20} show that for the linear case the CLR test is equivalent to the
continuously updating generalized method of moments (CU--GMM) test. \citet*{AndrewsMikusheva16} extend the CLR test
to address hypothesis testing in general moment condition models,
accommodating nonlinearity and HAC errors without imposing assumptions about
identification. However, implementing the
conditional likelihood ratio test requires finding the minimum of the
CU--GMM objective function. This function is non-convex, and, as shown by \Citet*{%
MoreiraNeweySharifvaghefi24}, the quasi-Newton methods that are commonly applied often fail
to reach the global minimum value in linear IV settings.

It is natural and entirely reasonable to rely on a discrete approximation of the
CU--GMM objective function to approximate
the minimum value, see for example the online supplement of \cite{AndrewsMikusheva16}. While the non-convexity of CU--GMM is well known, less is known about how discrete approximations affect the conditional distribution, size, and power of CLR-type tests. One might expect that, while the optimization is only approximate, the resulting approximation error would average out and therefore have little effect on the statistical properties of the implemented CLR test.
In this paper, we examine the consequences of such discrete optimization
schemes and show that discretization may actually distort test size
and diminish power with feasible grids (when the researcher does not use the
information on the true value $\theta $), or may spuriously inflate power with
infeasible grids (when the true unknown $\theta $ is artificially included
in the search). Importantly, similarity failure and power distortions can occur even when one moves only modestly away from the specific designs for which the grid was originally tailored. This reflects the lack of uniform guarantees over a class of data-generating processes that is inherent in the discretization procedure itself, rather than any deficiency of a particular grid.

A common intuition is that increasing the number of grid points should eliminate approximation error. While finer grids may improve performance in particular designs, as we show formally, increasing the number of grid points at a polynomial rate in the sample size $T$ does not guarantee that the resulting test preserves the similarity and power properties of the CLR test uniformly over a class of designs. Consequently, size and power distortions may persist even when the grid is refined at a polynomial rate.

Our results should not be interpreted as showing that every grid-based implementation performs poorly. Certain grids may perform well for particular designs. Rather, a central message of this paper is that grid-based implementations do not guarantee uniformly accurate approximations to the CLR test. We therefore underscore both the practicality of discretization-based approaches and the importance of developing alternative computational methods. For example, \citet{MoreiraNeweySharifvaghefi24} show that, in linear IV settings, the CU--GMM objective function can be rewritten as a ratio of polynomials. Consequently, the optimization problem can be reformulated as finding the roots of a polynomial equation, allowing existing computational algebra methods to compute the global minimum. This provides an implementation of the CLR test that preserves its theoretical properties.

Section \ref{sec:setup_and_clr_test} presents the general moment conditions
setup and explains the conditional likelihood ratio test. In Section \ref%
{sec:lr_stat_non_convex}, we explain the approximated CLR test based on the
discrete optimization algorithm and
establish its properties. Specifically, we provide sufficient conditions
under which the approximated CLR test has asymptotically zero size and
power. Finally, Section \ref{sec:MC-CLR} presents simulation studies to
illustrate the size and power issues that attend the use of the discrete optimization
algorithm to implement the CLR test in practice.

\bigskip

\noindent \textbf{Notations:} Generic finite positive constants are denoted
by $C_{i}$ for $i=1,2,\cdots $. $\lVert \mathbf{A}\rVert $ and $\lVert 
\mathbf{A}\rVert _{F}$ denote the spectral and Frobenius norms of matrix $%
\mathbf{A}$, respectively. $\left\Vert \mathbf{x}\right\Vert $ denotes the $%
\ell _{2}$ norm of vector $\mathbf{x}$. If $\{f_{n}\}_{n=1}^{\infty }$ and $%
\{g_{n}\}_{n=1}^{\infty }$ are both positive sequences of real numbers, then
we say $f_{n}=\ominus (g_{n})$ if there exist $n_{0}\geq 1$ and positive
constants $C_{0}$ and $C_{1}$, such that $\inf_{n\geq n_{0}}\left(
f_{n}/g_{n}\right) \geq C_{0}$ and $\sup_{n\geq n_{0}}\left(
f_{n}/g_{n}\right) \leq C_{1}$. Similarly, if $f_{iT}$\ and $g_{iT}$\ are
positive double sequences of real numbers for $i=1,2,3,\cdots $; and $%
T=1,2,3,\cdots $, then $f_{iT}=\ominus (g_{iT})$\ if there exist $T_{0}\geq
1 $ and positive constants $C_{0}$\ and $C_{1}$, such that $\inf_{T\geq
T_{0}}\left( f_{iT}/g_{iT}\right) \geq C_{0}$\ and $\sup_{T\geq T_{0}}\left(
f_{iT}/g_{iT}\right) \leq C_{1}$.

\section{Hypothesis testing based on moment conditions} \label{sec:setup_and_clr_test}

\subsection{Setup and Notation}

\label{subsec:setup}

Let $\varphi \left( X_{t},\theta \right) $ be a $k\times 1$ dimensional
function of the data $X_{t}$, where $\theta \in \mathbb{R}^{q}$ is a $%
q\times 1$ parameter. At the true parameter value $\theta ^{\ast }$, the
moments are equal to zero: 
\begin{equation}
\mathbb{E}\left[ \varphi \left( X_{t},\theta ^{\ast }\right) \right] =0.
\end{equation}%
Define the corresponding sample moment conditions as 
\begin{equation}
g_{T}(\theta )=T^{-1/2}\sum_{t=1}^{T}\varphi \left( X_{t},\theta \right) ,
\label{eq:general_sample_moment_condition}
\end{equation}%
and define the standardized moment as a function of $\theta $:%
\begin{equation}
m_{T}(\theta )=\mathbb{E}\left[ g_{T}(\theta )\right] =T^{-1/2}\sum_{t=1}^{T}%
\mathbb{E}\left[ \varphi \left( X_{t},\theta \right) \right] .
\end{equation}

Under mild regularity conditions (see, for example, \citet*{%
vanderVaartWellner96} for independent and identically distributed (i.i.d.)
data and for time series), empirical process
theory implies that 
\begin{equation}
g_{T}(\theta )-m_{T}(\theta )\overset{d}{=}G(\theta )+r_{T}(\theta ),
\label{eq:empirical-process}
\end{equation}%
where $G(\theta )$ is a mean-zero Gaussian process with a consistently
estimable variance matrix:%
\begin{equation}
\Sigma \left( \theta ,\tilde{\theta}\right) =\mathbb{E}\left[ G(\theta )G(%
\tilde{\theta})^{\prime }\right] ,
\end{equation}%
and $r_{T}(\theta )$ is a residual term that is asymptotically negligible.
For simplicity's sake, we assume $r_{T}(\theta )=0$ and that the
variance matrix $\Sigma (\theta ,\tilde{\theta})$ is known.

\begin{assumption}[Gaussian Process]
\label{assumption: Gaussian Process} 
\begin{equation}
g_{T}(\theta) - m_{T}(\theta) \sim \mathcal{N}(0, \Sigma(\theta, \theta)),
\quad \text{for all } \theta \in \mathbb{R}^{q},
\end{equation}
where $\Sigma(\theta, \theta)$ is known.
\end{assumption}

This assumption provides the Gaussian approximation that underpins the
conditional inference. It ensures quadratic forms in the moments behave like
likelihood ratio statistics, and makes the sufficient statistic construction
valid. This assumption is very general, but it may sometimes break down. For
instance, heavy tails or inconsistent covariance estimation would undermine
the conditional critical value function (CVF). In such cases, conditional pivotality may be lost: the
distribution of the LR statistic could depend on nuisance paths.

In the weak identification GMM literature, it is common to assume that the
vector of sample moment conditions is asymptotically normal. This normality
assumption is crucial for establishing that the CLR test (and related tests)
controls size at the desired level asymptotically. However, our results on
the failure of the approximated CLR test to achieve proper size and power
hold under weaker assumptions. In particular, using concentration
inequalities for sums of random variables with exponentially decaying
probability tails (which need not be normally distributed), the approximated likelihood-ratio-based test may fail to control size
under the null hypothesis and may have very low power under the alternative.

We are interested in testing the null hypothesis 
\begin{equation}
H_{0}:m_{T}(\theta _{0})=0\quad \text{ vs. }\quad H_{a}:m_{T}(\theta
_{0})\neq 0
\end{equation}%
where $\theta_{0}$ is a particular value of interest for $\theta$, using the CLR test discussed in Section \ref%
{subsec:CLR}.

A leading example of hypothesis testing based on moment conditions is the
linear IV model, as provided in Example \ref{ex:linear_iv}.

\begin{example}
\label{ex:linear_iv} Consider a linear IV model. Let $y\in \mathbb{R}^{T}$
denote the vector of observations on the dependent variable, and let $D\in 
\mathbb{R}^{T\times q}$ represent the matrix of observations on $q$
potentially endogenous regressors. Suppose the structural equation is given
by 
\begin{equation}
y=D\theta ^{\ast }+u,
\end{equation}%
where $u\in \mathbb{R}^{T}$ is the vector of structural error terms. Assume
that a matrix of instrumental variables $Z\in \mathbb{R}^{T\times k}$
satisfies the exclusion restriction 
\begin{equation}
\mathbb{E}\left( Z^{\prime }u\right) =0.
\end{equation}%
The corresponding sample moment conditions are defined as 
\begin{equation}
g_{T}(\theta )=\frac{1}{\sqrt{T}}Z^{\prime }\left( y-D\theta \right) =\left[
b(\theta )^{\prime }\otimes I_{k}\right] \mathrm{vec}\left( \frac{Z^{\prime
}Y}{\sqrt{T}}\right) ,  \label{eq:linear_iv_sample_moment_conditions}
\end{equation}%
where $b(\theta )=(1,-\theta ^{\prime })^{\prime }$ and $Y=(y,D)$. The
corresponding population moment conditions can thus be expressed as 
\begin{equation}
m_{T}(\theta )=\frac{1}{\sqrt{T}}\mathbb{E}\left[ Z^{\prime }(y-D\theta )%
\right] .
\end{equation}%
By substituting the structural equation for $y$, we have 
\begin{equation}
m_{T}(\theta )=\Gamma _{T}\left[ \sqrt{T}(\theta ^{\ast }-\theta )\right] 
\end{equation}%
where $\Gamma _{T}=\mathbb{E}(Z^{\prime }D/T)$. Finally, the
variance-covariance matrix $\Sigma \left( \theta ,\tilde{\theta}\right) $
can be expressed as 
\begin{equation}
\Sigma \left( \theta ,\tilde{\theta}\right) =\left[ b(\theta )^{\prime
}\otimes I_{k}\right] \widetilde{\Sigma }\left[ b(\tilde{\theta})\otimes
I_{k}\right] 
\end{equation}%
where $\widetilde{\Sigma }=\mathrm{var}\left( \mathrm{vec}\left( \frac{%
Z^{\prime }Y}{\sqrt{T}}\right) \right) $. In this setup, we are interested
in testing the null hypothesis $\theta ^{\ast }=\theta _{0}$ versus the alternative hypothesis $\theta ^{\ast
}\neq \theta _{0}$.
\end{example}

\subsection{Conditional likelihood ratio test}

\label{subsec:CLR}

Following \citet*{AndrewsMikusheva16}, we express the likelihood ratio
statistic as 
\begin{equation}  \label{eq:LR-def}
LR_{T} = Q_{T}\left(\theta_{0}\right) - \inf_{\theta \in \mathbb{R}^{q}}
Q_{T}(\theta)
\end{equation}
where 
\begin{equation}  \label{eq:cu-gmm}
Q_{T}(\theta) = g_{T}(\theta)^\prime\, \Sigma\left(\theta,
\theta\right)^{-1}\, g_{T}(\theta)
\end{equation}
is the continuously updating GMM (CU--GMM) objective function.

The critical value function for the likelihood ratio test is defined by 
\begin{equation}
c_{\alpha }=\min \left\{ c:P_{H_{0}}(LR_{T}>c)\leq \alpha \right\} 
\label{eq:cvf}
\end{equation}%
where $P_{H_{0}}(.)$ represents the probability measure under the null
hypothesis. As (\ref{eq:cvf}) makes clear, the critical value function
for the likelihood ratio test depends on the distribution of the statistic $%
LR_{T}$ under the null hypothesis, which in turn depends on the (generally
unknown) function $m_{T}(\theta )$. However, as \cite%
{AndrewsMikusheva16} discuss, once we condition on 
\begin{equation}
h_{T}(\theta )=g_{T}(\theta )-\Sigma \left( \theta ,\theta _{0}\right)
\Sigma \left( \theta _{0},\theta _{0}\right) ^{-1}\,g_{T}(\theta _{0}),
\label{eq:sufficient_stat}
\end{equation}%
the distribution of $LR_{T}$ under the null hypothesis does not depend on $%
m_{T}(\theta )$.

To see why this is the case, we can set $h_{T}(\theta )=\bar{h}%
_{T}(\theta )$ and write 
\begin{equation}
g_{T}(\theta )=\bar{h}_{T}(\theta )+\Sigma \left( \theta ,\theta _{0}\right)
\Sigma \left( \theta _{0},\theta _{0}\right) ^{-1}\,g_{T}(\theta _{0}).
\label{eq:g_conditional_h}
\end{equation}%
From (\ref{eq:sufficient_stat}), we obtain%
\begin{eqnarray}
Cov\left( g_{T}(\tilde{\theta}),h_{T}\left( \theta \right) \right) &=&%
\mathbb{E}\left[ \left( g_{T}\left( \tilde{\theta}\right) -m_{T}\left( 
\tilde{\theta}\right) \right) h_{T}\left( \theta \right) ^{\prime }\right] \\
&=&\Sigma \left( \tilde{\theta},\theta \right) -\Sigma \left( \tilde{\theta}%
,\theta _{0}\right) \Sigma \left( \theta _{0},\theta _{0}\right) ^{-1}\Sigma
\left( \theta _{0},\theta \right) .  \notag
\end{eqnarray}%
In particular, for all $\theta $, 
\begin{equation}
Cov\left( g_{T}(\theta _{0}),h_{T}\left( \theta \right) \right) =0.
\end{equation}

Since $g_{T}(\theta _{0})$ and $h_{T}(\theta )$ follow a joint normal
distribution, it follows that $g_{T}(\theta _{0})$ is independent of $%
h_{T}(\theta )$. Consequently, from (\ref{eq:g_conditional_h}), the distribution of $g_{T}(\theta )$ conditional on $h_{T}(\theta )=%
\bar{h}_{T}(\theta )$ depends on $m_{T}(\theta )$ only at $\theta =\theta
_{0}$. Specifically, 
\begin{eqnarray}
&&\left[ g_{T}(\theta )-h_{T}(\theta )\right] |\left\{ h_{T}(\theta )=\bar{h}%
_{T}(\theta )\right\} \\
&\sim &\mathcal{N}\left( \Sigma (\theta ,\theta _{0})\Sigma (\theta
_{0},\theta _{0})^{-1}m_{T}(\theta _{0}),\Sigma (\theta ,\theta _{0})\Sigma
(\theta _{0},\theta _{0})^{-1}\Sigma (\theta _{0},\theta )\right) .  \notag
\end{eqnarray}%
Under the null hypothesis, $m_{T}\left( \theta _{0}\right) =0$,%
\begin{equation}
\left[ g_{T}(\theta )-h_{T}(\theta )\right] |\left\{ h_{T}(\theta )=\bar{h}%
_{T}(\theta )\right\} \sim \mathcal{N}\left( 0,\Sigma (\theta ,\theta
_{0})\,\Sigma (\theta _{0},\theta _{0})^{-1}\,\Sigma (\theta _{0},\theta
)\right) .
\end{equation}%
Therefore, under the null hypothesis, the distribution of $LR_{T}$
conditional on $h_{T}(\theta )=\bar{h}_{T}(\theta )$ does not depend on $%
m_{T}(\theta )$. This allows us to construct the conditional critical value
function $c_{\alpha }(\bar{h}_{T})$, defined by 
\begin{equation}
c_{\alpha }(\bar{h}_{T})=\min \left\{ c:\Pr_{H_{0}}\left( \left.
LR_{T}>c\right\vert h_{T}=\bar{h}_{T}\right) \leq \alpha \right\} .
\label{eq:conditional_cvf}
\end{equation}%
The CLR test then rejects the null hypothesis when the likelihood ratio
statistic, $LR_{T}$, is greater than the conditional critical value
function, $c_{\alpha }(\bar{h}_{T})$.

\section{Non-convex optimization and discrete approximation}

\label{sec:lr_stat_non_convex}

While \citet*{AndrewsMikusheva16} show that the CLR test is asymptotically
similar, we require a reliable optimization procedure to
ensure that its theoretical properties do indeed hold in practice. To implement the
CLR test, we need to find the infimum of the CU--GMM objective function with
respect to $\theta \in \mathbb{R}^{q}$, which is generally non-convex. 
In this section, we consider implementations based on a discrete approximation
to $\inf_{\theta \in \mathbb{R}^{q}} Q_T(\theta)$. We describe the resulting
approximated CLR test within the general framework of Section
\ref{subsec:setup} and establish the conditions
under which the approximated CLR test fails to be similar and has zero
power.

Consider the set 
\begin{equation}
\Theta _{P_{T}}=\{\theta _{0},\theta _{1},\dots ,\theta _{P_{T}}\}\subset 
\mathbb{R}^{q}
\end{equation}%
as a generic search set for the discrete approximation method. The
approximated likelihood ratio statistic is defined as 
\begin{equation}
\widetilde{LR}_{T}=Q_{T}(\theta _{0})-\min_{\theta \in \Theta
_{P_{T}}}Q_{T}(\theta )  \label{eq:approx_lr_stat}
\end{equation}%
where $Q_{T}(\theta )$ is the CU--GMM objective function given by (\ref%
{eq:cu-gmm}). The conditional critical value function based on the
approximated likelihood ratio statistic, denoted by $\tilde{c}_{\alpha }(%
\bar{h}_{T})$, is 
\begin{equation}
\tilde{c}_{\alpha }(\bar{h}_{T})=\min \left\{ c:\Pr_{H_{0}}\left( \left. 
\widetilde{LR}_{T}>c\right\vert h_{T}=\bar{h}_{T}\right) \leq \alpha
\right\} .  \label{eq:cvf_approx_clr}
\end{equation}%
Finally, following \citet*{AndrewsMikusheva16}, the approximated CLR test
rejects the null hypothesis when $\widetilde{LR}_{T}>\tilde{c}_{\alpha
}(h_{T})$.

From this point on, we assume the search set for the approximated
likelihood ratio test satisfies the following properties:

\begin{assumption}
\label{assumption: search_set} (i) The search set, $\Theta_{P_{T}}$, is
deterministic. (ii) It contains the value of $\theta$ under the null
hypothesis, denoted by $\theta_{0}$. (iii) The number of elements in the
search set, $P_{T}$, can increase at a polynomial rate with $T$.
\end{assumption}

We assume that the search set $\Theta_{P_T}$ is deterministic to simplify the analysis. Deterministic search sets also arise in practice. For example, \citet*{AndrewsMikusheva16} employ a deterministic grid in their simulation study of the CLR test for linear IV models with HAC errors. The
analysis of the CLR test's behavior based on the
approximated LR statistic could be extended to cases where $\Theta _{P_{T}}$
is random, provided that $m_{T}(\theta )$ is not in the local neighborhood
of zero for all $\theta \in \Theta _{P_{T}}\backslash \{\theta _{0}\}$ with
high probability. Including the value of $\theta $ in the search set under the null
hypothesis, $\theta _{0}$, ensures that the approximated
LR statistic is non-negative.

Consider the constrained estimator 
\begin{equation}
\tilde{\theta}=\arg \min_{\theta \in \Theta _{P_{T}}\backslash \{\bar{\theta}%
\}}Q_{T}\left( \theta \right)
\end{equation}%
for an arbitrary value $\bar{\theta}$. Then, obtain the following bound:%
\begin{eqnarray}
\Pr \left[ Q_{T}\left( \tilde{\theta}\right) -Q_{T}\left( \bar{\theta}%
\right) >c_{0}\right] &=&\Pr \left( \cap _{i=1}^{P_{T}}\left\{ Q_{T}\left(
\theta _{i}\right) -Q_{T}\left( \bar{\theta}\right) >c_{0}\right\} \right) \\
&=&1-\Pr \left( \cup _{i=1}^{P_{T}}\left\{ Q_{T}\left( \bar{\theta}\right)
-Q_{T}\left( \theta _{i}\right) \geq -c_{0}\right\} \right)  \notag \\
&\geq &1-\sum_{i=1}^{P_{T}}\Pr \left( Q_{T}\left( \bar{\theta}\right)
-Q_{T}\left( \theta _{i}\right) \geq -c_{0}\right) .  \notag
\end{eqnarray}

The problem with the discrete approximation method remains even if the
number of elements in the search set increases at a polynomial rate with $T$%
. Suppose that among the values of $\theta $ within the grid search, the
population value of the sample moment conditions evaluated at $\theta _{0}$, 
$m(\theta _{0})$, is the closest to $m(\theta ^{\ast })=0$, and there is no
other point in the local neighborhood of $\theta _{0}$. By Assumption \ref%
{assumption: Gaussian Process}, $g(\theta )$ is Gaussian, and so, has an
exponentially decaying probability tail. That is, the probability of the event in
which the value of the CU--GMM objective function, evaluated at an arbitrary
point in the search set, exceeds the value of the CU--GMM objective function,
evaluated at $\theta _{0}$, decreases to zero at an exponential rate.
Therefore, the sum of these probabilities over all $\theta \in \Theta
_{P_{T}}\backslash \{\theta _{0}\}$ approaches zero, even while the number
of elements in the search set grows at a polynomial rate. If the polynomial
rate at which the number of grid points increases with $T$ is too slow, it
is still possible that no point in the neighborhood of $\theta _{0}$ will be
included in the grid search.

Although the true value of $\theta ^{\ast }$ is
unknown, the search set for the exact likelihood ratio statistic, which is $%
\mathbb{R}^{q}$, contains $\theta ^{\ast }$ and infinitely many points in
the local neighborhood of $m_{T}(\theta ^{\ast })=0$. However, it is
impossible to ensure that all points in the local neighborhood of $%
m_{T}(\theta ^{\ast })=0$ are included in the search set of the approximated
likelihood ratio statistic. Here, we establish that when none of the points in the local neighborhood of $m_{T}(\theta ^{\ast
})=0$ are included in the search set, apart from $\theta
_{0}$, the asymptotic distribution of the
approximated likelihood ratio statistic, conditional on $h(\theta )$,
differs significantly from the distribution of the exact conditional
likelihood ratio statistic. In particular, under the null hypothesis, the
conditional probability that $Q(\theta _{0})=\min_{\theta \in \Theta
_{P_{T}}}Q_{T}(\theta )$ given $h(\theta )$ approaches one as $T\rightarrow
\infty $. Consequently, as Proposition~\ref%
{prop:distribution_approx_lr} establishes, the conditional approximated likelihood ratio
statistic has an asymptotically degenerate distribution at zero under the
null hypothesis, where $m_{T}(\theta _{0})=0$.

\begin{proposition}
\label{prop:distribution_approx_lr} Suppose Assumptions \ref{assumption:
Gaussian Process} and \ref{assumption: search_set} hold. For a given value
of $h_{T}(\theta )$, denoted by $\bar{h}_{T}(\theta )$, assume that for all $%
\theta \in \Theta _{P_{T}}\backslash \{\theta _{0}\}$, $\left\Vert \Sigma
\left( \theta ,\theta \right) ^{-1/2}\bar{h}_{T}\left( \theta \right)
\right\Vert =\ominus (T^{\kappa -\frac{1}{2}})$ for some $1/2<\kappa \leq 1$%
. Then, under the null hypothesis, i.e.\ $m_{T}(\theta _{0})=0$, there exist 
$T_{0}$ and a positive constant $C$ such that for all $T\geq T_{0}$, 
\begin{equation}
\Pr \left( \widetilde{LR}_{T}=0|h_{T}(\theta )=\bar{h}_{T}(\theta )\right)
\geq 1-\exp \left( -CT^{(2\kappa -1)}\right) .
\end{equation}
\end{proposition}

Since, under the null hypothesis, $\widetilde{LR}_{T}$ has an asymptotic
degenerate conditional distribution at zero, we can conclude that there
exists a $T_{0}$ such that for all $T\geq T_{0}$, $\tilde{c}_{\alpha }(\bar{h%
}_{T})=0$. Proposition \ref{prop:cvf_behavior} below formalizes this
statement.

\begin{proposition}
\label{prop:cvf_behavior} Suppose Assumptions \ref{assumption: Gaussian
Process} and \ref{assumption: search_set} hold. For a given value of $%
h_{T}(\theta )$, denoted by $\bar{h}_{T}(\theta )$, assume that for all $%
\theta \in \Theta _{P_{T}}\backslash \{\theta _{0}\}$, $\left\Vert \Sigma
\left( \theta ,\theta \right) _{T}^{-1/2}\bar{h}\left( \theta \right)
\right\Vert =\ominus (T^{\kappa -\frac{1}{2}})$ for some $1/2<\kappa \leq 1$%
. Then there exists a $T_{0}$ such that for all $T\geq T_{0}$, we have $%
\tilde{c}_{\alpha }(\bar{h}_{T})=0$.
\end{proposition}

The approximated CLR statistic therefore degenerates,
collapsing to zero under the null. Consequently, the test becomes biased and
fails to achieve good power, even against local alternatives. The problem lies not with the conditional construction itself, but with
the discrete approximation used in implementation. Due to the continuity of
the (conditional) power function, the poor size and power properties of the
grid CLR test persist even under weaker conditions, such as $\kappa =1/2$.
In this case, some terms do not vanish and the CLR manages to avoid degenerating.
Nevertheless, the undesirable size and power behavior persists under
these more general conditions, as we shall see in the simulations.

The strategy of the proof is to compare the CU--GMM objective at $\theta _{0}$ and
at grid points. Under the null, the global minimizer lies near solutions to $%
m_{T}(\theta )=0$, but if the grid excludes that neighborhood, the minimum
is attained at $\theta _{0}$ with probability approaching one. Thus the LR
collapses to zero, as does the conditional critical value. Under
alternatives, the same reasoning applies: deterministic gaps away from $%
\theta ^{\ast }$ dominate stochastic variation, forcing the discrete minimum
to pick $\theta _{0}$ at the expense of power.

Propositions \ref{prop:distribution_approx_lr} and \ref{prop:cvf_behavior}
rely on $\left\Vert \Sigma \left( \theta ,\theta \right) ^{-1/2}\bar{h}%
_{T}\left( \theta \right) \right\Vert =\ominus (T^{\kappa -\frac{1}{2}})$
for all $\theta \in \Theta _{P_{T}}\setminus \{\theta _{0}\}$, where $%
1/2<\kappa \leq 1$ is a constant. Next, we see the conditions
that ensure that this rate holds with probability approaching one. When $\theta
\neq \theta _{0}$, by Assumption \ref{assumption: Gaussian Process}, we have 
\begin{equation}
\left[ h_{T}(\theta )-\mu _{T,h}(\theta )\right] \sim \mathcal{N}\left(
0,V_{h}(\theta ,\theta _{0})\right) 
\end{equation}%
where 
\begin{equation}
\mu _{T,h}(\theta )=m_{T}(\theta )-\Sigma (\theta ,\theta _{0})\Sigma
(\theta _{0},\theta _{0})^{-1}m_{T}(\theta _{0})
\end{equation}%
and 
\begin{equation}
V_{h}(\theta ,\theta _{0})=\Sigma (\theta ,\theta )-\Sigma (\theta ,\theta
_{0})\Sigma (\theta _{0},\theta _{0})^{-1}\Sigma (\theta _{0},\theta ).
\end{equation}%
Therefore, 
\begin{equation}
\Sigma \left( \theta ,\theta \right) ^{-1/2}\left[ h_{T}(\theta )-\mu
_{T,h}(\theta )\right] \sim \mathcal{N}\left( 0,\widetilde{V}_{h}(\theta
,\theta _{0})\right) 
\end{equation}%
where 
\begin{equation}
\widetilde{V}_{h}(\theta ,\theta _{0})=I_{k}-\Sigma \left( \theta ,\theta
\right) ^{-1/2}\Sigma (\theta ,\theta _{0})\Sigma (\theta _{0},\theta
_{0})^{-1}\Sigma (\theta _{0},\theta )\Sigma \left( \theta ,\theta \right)
^{-1/2}.
\end{equation}

Under the null hypothesis, $m_{T}(\theta _{0})=0$ and hence $%
\mu _{T,h}(\theta )=m_{T}(\theta )$. Assumption \ref{assumption:
noncentrality size} below establishes how both upper and lower bounds of the
noncentrality parameter $\left\Vert \Sigma \left( \theta ,\theta \right)
^{-1/2}m_{T}\left( \theta \right) \right\Vert $ grow with $T$. An increase in the rate of divergence affects the exponential
rate of decay in the conditional probability that the approximated
likelihood ratio is not exactly equal to zero. Otherwise, the size and power distortions of the approximated CLR test, documented below, continue to hold as long as $\left\Vert \Sigma \left( \theta ,\theta \right)
^{-1/2}m_{T}\left( \theta \right) \right\Vert $ diverges to infinity as $%
T\rightarrow \infty $. These rates for the noncentrality parameter
guarantee the same rates of growth for $\left\Vert \Sigma \left( \theta ,\theta
\right) ^{-1/2}\bar{h}_{T}\left( \theta \right) \right\Vert $ with
probability approaching one as $T\rightarrow \infty $.

\begin{assumption}[Noncentrality]
\label{assumption: noncentrality size} For all $\theta \in \Theta
_{P_{T}}\backslash \{\theta _{0}\}$, $\left\Vert \Sigma \left( \theta
,\theta \right) ^{-1/2}m_{T}\left( \theta \right) \right\Vert =\ominus
(T^{\kappa -\frac{1}{2}})$ for some $1/2<\kappa \leq 1$.
\end{assumption}

Building on Propositions \ref{prop:distribution_approx_lr} and \ref%
{prop:cvf_behavior}, Theorem \ref{thm:
size_approx_clr_test} establishes that the approximated CLR test that rejects
the null hypothesis when $\widetilde{LR}_{T}>\tilde{c}_{\alpha }(h_{T})$,
has asymptotically zero size.

\begin{theorem}
\label{thm: size_approx_clr_test} Suppose Assumptions \ref{assumption:
Gaussian Process}, \ref{assumption: search_set}, and \ref{assumption:
noncentrality size} hold. Then, under the null hypothesis $%
m_{T}(\theta _{0})=0$, there exist $T_{0}$ and a positive constant $C$ such
that for all $T\geq T_{0}$, 
\begin{equation}
\Pr \left( \widetilde{LR}_{T}>\tilde{c}_{\alpha }(h_{T})\right) \leq \exp
(-CT^{(2\kappa -1)}).
\end{equation}
\end{theorem}

Altering the rejection rule in any way will not resolve this
problem. If we modify the rule to reject the null whenever $\widetilde{LR}%
_{T}\geq \tilde{c}_{\alpha }(h_{T})$, then by Proposition~\ref%
{prop:cvf_behavior} the conditional critical value satisfies $\tilde{c}%
_{\alpha }(h_{T})=0$ with probability approaching one as $T\rightarrow
\infty $. Since $LR_{T}\geq 0$ by construction, this implies 
\begin{equation}
\Pr_{H_{0}}\left( \widetilde{LR}_{T}\geq \tilde{c}_{\alpha }(h_{T})\right) =1
\end{equation}%
for all large $T$, so the test has asymptotic size equal to one.
Taking a different approach, we could attempt to randomize at the boundary case $%
LR_{T}=c_{\alpha }(h_{T})$, rejecting with probability $\alpha $. Under $%
H_{0}$, we have $\Pr (\widetilde{LR}_{T}=\tilde{c}_{\alpha
}(h_{T}))\rightarrow 1$ and $\Pr (\widetilde{LR}_{T}>\tilde{c}_{\alpha
}(h_{T}))\rightarrow 0$, so the limiting size of this randomized test equals 
$\alpha $. We would effectively be flipping a coin at the boundary. This,
however, introduces purely random rejections under the null and yields no
gain in power. In fact, under the alternatives described in Theorem~\ref%
{thm: power_approx_clr_test}, where the discrete search fails to capture the
neighborhood of $m_{T}(\theta ^{\ast })=0$, we also have $\Pr (\widetilde{LR}%
_{T}=\tilde{c}_{\alpha }(h_{T}))\rightarrow 1$, so the power of such a
randomized test collapses to $\alpha $.

Thus, whether we reject the null when $\widetilde{LR}_{T}$ is strictly
larger than the conditional quantile (leading to asymptotic zero null
rejection probabilities), larger or equal (leading to asymptotic size one),
or randomization at equality (leading to unnecessary power loss), the
outcome is unsatisfactory. The fundamental issue is the discrete
approximation of the CU--GMM objective function. A reliable implementation must guarantee that the global infimum of the non-convex CU--GMM objective function is attained.

In the linear IV case, this can be accomplished by the polynomial method of %
\citet*{MoreiraNeweySharifvaghefi24}. This polynomial method guarantees the global minimum,
thereby reconciling the implementation with the theoretical properties of
the CLR test. \ We can examine the conditions under which, for
all $\theta \in \Theta _{P_{T}}\backslash \{\theta _{0}\}$, 
\begin{equation}
\left\Vert \Sigma \left( \theta ,\theta \right) ^{-1/2}m_{T}\left( \theta
\right) \right\Vert =\ominus (T^{\kappa -\frac{1}{2}})
\end{equation}%
in the linear IV models provided in Example \ref{ex:linear_iv}, and thus,
based on Theorem \ref{thm: size_approx_clr_test}, the approximated CLR test
has zero size. In these models, we express $m_{T}(\theta )$ as 
\begin{equation}
m_{T}(\theta )=\Gamma _{T}\left[ \sqrt{T}\left( \theta ^{\ast }-\theta
\right) \right] .
\end{equation}%
The following assumption establishes conditions for strong identification.

\begin{assumption}[Strong\ Instruments]
\label{assumption: SIV} (i) $\Gamma _{T}$ converges to a full column rank
matrix; and (ii) for all $\theta \in \Theta _{P_{T}}\backslash \{\theta
_{0}\}$, $\left\Vert \theta -\theta ^{\ast }\right\Vert =\ominus (T^{\kappa
-1})$, where $\frac{1}{2}<\kappa \leq 1$ is a constant.
\end{assumption}

Under Assumption \ref{assumption: SIV}, $\left\Vert m_{T}(\theta
)\right\Vert =\ominus (T^{\kappa -\frac{1}{2}})$, and so, Assumption \ref%
{assumption: noncentrality size} holds. Thus, in linear models with strong
IV, the approximated CLR test has asymptotic zero size under the null
hypothesis, as formally stated in Corollary \ref{corollary: strong iv null
hypothesis}.

\begin{corollary}
\label{corollary: strong iv null hypothesis} Consider the linear IV models
with strong instrumental variables discussed in Example \ref{ex:linear_iv}.
Suppose Assumptions \ref{assumption: Gaussian Process}, \ref{assumption:
search_set}, and \ref{assumption: SIV} hold. Then, under the null hypothesis, $m_{T}(\theta _{0})=0$, there exist $T_{0}$ and a positive constant $C$
such that for all $T\geq T_{0}$, 
\begin{equation}
\Pr \left( \widetilde{LR}_{T}>\tilde{c}_{\alpha }(h_{T})\right) \leq \exp
(-CT^{2\kappa -1}).
\end{equation}
\end{corollary}

Under alternative hypotheses, $m_{T}(\theta _{0})\neq 0$. We replace
Assumption \ref{assumption: noncentrality size} by the following assumption.

\begin{assumption}[Noncentrality Difference]
\label{assumption: noncentrality power} For all $\theta \in \Theta
_{P_{T}}\backslash \{\theta _{0}\}$, 
\begin{equation}
d_{T}\left( \theta ,\theta _{0}\right) =\left\Vert \Sigma \left( \theta
,\theta \right) ^{-1/2}m_{T}\left( \theta \right) \right\Vert
^{2}-\left\Vert \Sigma \left( \theta _{0},\theta _{0}\right)
^{-1/2}m_{T}\left( \theta _{0}\right) \right\Vert ^{2}=\ominus (T^{2\kappa
-1}),
\end{equation}%
for some $1/2<\kappa \leq 1$.
\end{assumption}

If $m_{T}(\theta _{0})$ is relatively closer to $m_{T}(\theta ^{\ast })=0$
compared to other $m_{T}(\theta )$ for $\theta \in \Theta _{P_{T}}\backslash
\{\theta _{0}\}$, then we can further establish that the approximated
likelihood ratio statistic has an asymptotically degenerate conditional
distribution at zero. Consequently, the approximated CLR test has asymptotic
zero power, as formally stated in Theorem \ref{thm: power_approx_clr_test}.

\begin{theorem}
\label{thm: power_approx_clr_test} Suppose Assumptions \ref{assumption:
Gaussian Process}, \ref{assumption: search_set}, and \ref{assumption:
noncentrality power} hold. Then, under alternative hypotheses $%
m_{T}(\theta _{0})\neq 0$, there exist $T_{0}$ and a positive constant $C$
such that for all $T\geq T_{0}$ 
\begin{equation}
\Pr \left( \widetilde{LR}_{T}>\tilde{c}_{\alpha }(h_{T})\right) \leq \exp
(-CT^{(2\kappa -1)}).
\end{equation}
\end{theorem}

Theorems \ref{thm: size_approx_clr_test} and \ref{thm: power_approx_clr_test} show that deterministic discretization methods with polynomially growing search sets do not provide uniform guarantees that the approximated CLR test preserves the similarity and power properties of the exact CLR test.

A prominent example of Assumption \ref{assumption: noncentrality power} is
testing for local alternative hypotheses in the linear IV models described
in Example~\ref{ex:linear_iv}, where $\theta _{0}\neq \theta ^{\ast }$ but
is nevertheless located in a local neighborhood of $\theta ^{\ast }$ such that $%
\left\Vert \theta ^{\ast }-\theta _{0}\right\Vert =\ominus (1/\sqrt{T})$.
When the instrumental variables are strong, under local alternatives, we
have $\left\Vert m_{T}\left( \theta _{0}\right) \right\Vert =\ominus (1)$.
Assumption \ref{assumption: noncentrality power} holds if no other values of 
$\theta $ in the search set lie within the neighborhood of $\theta
^{\ast }$, such that $\left\Vert m_{T}\left( \theta \right) \right\Vert
=\ominus \left( T^{\kappa -\frac{1}{2}}\right) $ for some $1/2<\kappa \leq 1$%
. Therefore, Theorem \ref{thm: power_approx_clr_test} implies that the
approximated CLR test has asymptotically zero power (see Corollary \ref%
{corollary: strong iv local alternatives}).

\begin{corollary}
\label{corollary: strong iv local alternatives} Consider the linear IV
models with strong instrumental variables discussed in Example \ref%
{ex:linear_iv}. Suppose Assumptions \ref{assumption: Gaussian Process}, \ref%
{assumption: search_set}, and \ref{assumption: SIV} hold. Then, under the local
alternative hypothesis $m_{T}(\theta _{0})=\ominus (1)$,
there exist $T_{0}$ and a positive constant $C$ such that for all $T\geq
T_{0}$, 
\begin{equation}
\Pr \left( \widetilde{LR}_{T}>\tilde{c}_{\alpha }(h_{T})\right) \leq \exp
(-CT^{2\kappa -1}).
\end{equation}
\end{corollary}

By including $\theta _{0}$ in the search set $\Theta _{P_{T}}$, we ensure
that the true parameter value $\theta ^{\ast }$ is included in the search
set under the null hypothesis. Moreover, under local alternatives (even
though $\theta ^{\ast }\neq \theta _{0}$), we have 
\begin{equation}
\Vert m_{T}(\theta _{0})-m_{T}(\theta ^{\ast })\Vert =\Vert m_{T}(\theta
_{0})\Vert =\ominus \left( 1\right) ,
\end{equation}%
which implies that 
\begin{equation}
\min_{\theta \in \Theta _{P_{T}}}Q_{T}(\theta )-\inf_{\theta \in \mathbb{R}%
^{k}}Q_{T}(\theta )=O_{p}\left( 1\right) .
\end{equation}

Now consider an alternative hypothesis in which $\theta _{0}$ is not in the
local neighborhood of $\theta ^{\ast }$, such that 
\begin{equation}
\Vert m_{T}(\theta _{0})-m_{T}(\theta ^{\ast })\Vert =\Vert m_{T}(\theta
_{0})\Vert =\ominus \left( T^{\kappa -\frac{1}{2}}\right) 
\end{equation}%
where $1/2<\kappa \leq 1$ is a constant. In this scenario, consider the case
in which all other points in the search set are also far from $\theta ^{\ast
}$. Then, for all $\theta \in \Theta _{P_{T}}\backslash \{\theta _{0}\}$, we
have $d_{T}\left( \theta ,\theta _{0}\right) =\ominus \left( T^{\kappa -%
\frac{1}{2}}\right) $, and so, 
\begin{equation}
\min_{\theta \in \Theta _{P_{T}}}Q_{T}(\theta )-\inf_{\theta \in \mathbb{R}%
^{k}}Q_{T}(\theta )=\tilde{\eta}_{T}^{2}+O_{p}\left( \sqrt{T}\right) 
\end{equation}%
where $\tilde{\eta}_{T}=\ominus \left( T^{\kappa -\frac{1}{2}}\right) $.
Consequently, the difference between the exact LR statistic, $LR_{T}$, and
the approximated LR statistic, $\widetilde{LR}_{T}$, satisfies 
\begin{equation}
LR_{T}-\widetilde{LR}_{T}=\min_{\theta \in \Theta _{P_{T}}}Q_{T}(\theta
)-\inf_{\theta \in \mathbb{R}^{k}}Q_{T}(\theta )=\tilde{\eta}%
_{T}^{2}+O_{p}\left( \sqrt{T}\right) ,
\end{equation}%
which diverges to infinity as $T\rightarrow \infty $. The discrepancy
between the exact and approximated likelihood ratio statistics can thus lead
to significant power deficiencies for the approximated CLR test. In
particular, if Assumption \ref{assumption: noncentrality power} holds, by
Theorem \ref{thm: power_approx_clr_test}, the approximated CLR test has zero
asymptotic power.

It is natural to consider the search set $\Theta _{P_{T}}^{\ast }$ which
artificially includes the unknown true parameter $\theta ^{\ast }$ (which
makes the approximated CLR test infeasible) for simulation purposes. The
infeasible $LR$ statistic would then be%
\begin{equation}
\widetilde{LR}_{T}^{\ast }=Q_{T}(\theta _{0})-\min_{\theta \in \Theta
_{P_{T}}^{\ast }}Q_{T}(\theta ).  \label{eq:approx_lr_stat infeasible}
\end{equation}%
This would ensure that the difference between the approximated and exact $LR$
statistics is stochastically bounded, namely: 
\begin{equation}
LR_{T}-\widetilde{LR}_{T}^{\ast }=O_{p}(1).
\label{eq:true and infeasible LR stat}
\end{equation}%
In contrast to the feasible likelihood ratio statistic, the difference between
the infeasible and the exact likelihood ratio statistics is bounded. This
fact might misleadingly suggest that artificially including $\theta ^{\ast }$
in the search set would cause negligible distortion. However, the
difference in (\ref{eq:true and infeasible LR stat}) does not vanish
asymptotically. It may lead to size or power distortions relative to the
exact (conditional) likelihood ratio test. Theorem \ref{thm:
power_infeasible_approx_clr_test} below formalizes this concern and its
effect on numerical properties of the infeasible grid CLR test. It
establishes that the infeasible version of the approximated CLR test that
includes $\theta ^{\ast }$ in the search set achieves asymptotic power of
one as $T\rightarrow \infty $. This contrasts with Theorem \ref{thm:
power_approx_clr_test}, which shows that the feasible version that excludes $\theta ^{\ast }$
from the search set may have power converging to zero as $T\rightarrow
\infty $. In the supplement, we also show distortions caused by implementing the
infeasible LR test.

\begin{theorem}
\label{thm: power_infeasible_approx_clr_test} Suppose Assumptions \ref%
{assumption: Gaussian Process} and \ref{assumption: search_set} hold. Then, under non-local alternatives
with $\left\Vert m_{T}\left( \theta _{0}\right) \right\Vert =\ominus
(T^{\kappa -\frac{1}{2}})$ and $1/2<\kappa \leq 1$, there exist $T_{0}$ and
a positive constant $C$ such that for all $T\geq T_{0}$, 
\begin{equation}
\Pr \left( \widetilde{LR}_{T}^{\ast }>\tilde{c}_{\alpha }^{\ast
}(h_{T})\right) \geq 1-\exp \left( -CT^{(2\kappa -1)}\right) .
\end{equation}
\end{theorem}

Since the conditional critical value function $\tilde{c}(h_{T})$ for the approximated CLR test is less than
or equal to the critical value function $c(h_{T})$ of the exact CLR test, and
given that $\tilde{c}(h_{T})=0$ with probability approaching one as $%
T\rightarrow \infty $ (by Proposition \ref{prop:cvf_behavior} and Lemma S-1.2, available in the supplement), the infeasible
approximated CLR test may exhibit artificially inflated power compared to
the exact CLR test. To illustrate this point, in Section \ref{sec:MC-CLR} we consider, as an example, the infeasible search set used by \citet*{AndrewsMikusheva16}, which includes the true parameter value $\theta^{\ast}$.

The negative results above should not be interpreted as implying that all grid-based approximations fail. In settings with a compact parameter space, a sufficiently smooth mean function, and a refined grid such that every point in the parameter space is $o(T^{-1/2})$ away from a grid point, one would expect the discretized optimization problem to approximate the global minimum of the CU--GMM objective function accurately. Under such conditions, the mechanism underlying Theorems \ref{thm: size_approx_clr_test} and \ref{thm: power_approx_clr_test} and the corresponding corollaries is ruled out.

Our results instead highlight a lack of uniformity across data-generating processes. The location and curvature of the CU--GMM objective function may vary substantially across realizations and parameter values. Consequently, a discretization scheme that performs well for one design need not perform well for another. Theorems \ref{thm: size_approx_clr_test} and \ref{thm: power_approx_clr_test} show that polynomial growth in the number of grid points, by itself, is not sufficient to guarantee that the resulting approximated CLR test preserves the similarity and power properties of the exact CLR test uniformly over a class of data-generating processes.

Similarly, adaptive refinement schemes or multi-start local optimization procedures may improve performance in practice by increasing the likelihood of locating regions near the global minimum. However, without guarantees that the global infimum is attained, such procedures do not rule out the mechanisms studied in this paper. Consequently, they do not, by themselves, ensure that the resulting CLR implementation preserves the similarity and power properties of the exact CLR test uniformly over a class of data-generating processes.

From a practical perspective, implementing a grid-based approximation still requires the researcher to choose features such as the grid boundaries, the initial grid density, or the refinement strategy. These choices depend on features of the objective function that are generally unknown ex ante, and there is typically little guidance regarding how fine the search must be in order to accurately approximate the global minimum of the CU--GMM objective function. For example, in the IES application considered below, it is not obvious whether a grid with 25, 400, or 4000 points is adequate, nor how the boundary points should be selected. The simulation evidence reported in Section \ref{sec:MC-CLR} illustrates both the potential success and the potential failure of such discretization schemes.

\section{Simulations\label{sec:MC-CLR}}

\subsection{Design with homoskedastic errors}

To illustrate the role of accuracy in the computation of the CU--GMM
estimator for the performance of the CLR test, we consider a simple linear
IV regression with homoskedastic and uncorrelated errors. The variance of $%
\text{vec}\!\left( \tfrac{Z^{\prime }Y}{\sqrt{T}}\right) $, denoted by $%
\widetilde{\Sigma }$, can be written as $\widetilde{\Sigma }=\Omega \otimes
Z^{\prime }Z/T$, where $\Omega $ is the $2\times 2$ variance matrix of $%
(y_{i},D_{i})^{\prime }$ conditional on $Z$. Consequently, 
\begin{equation}
Q(\theta )=\frac{b(\theta )^{\prime }Y^{\prime }P_{Z}Yb(\theta )}{b(\theta
)^{\prime }\Omega b(\theta )},  \label{eq:hom_cu_gmm}
\end{equation}%
so the CU--GMM objective function takes the form of a generalized Rayleigh
quotient, and minimizing $Q(\theta )$ is equivalent to solving a generalized
eigenvalue problem.

We begin with the homoskedastic case to show that distortions in size and
power can arise even in this simple setting. We follow %
\citet{ChaoHasmanNeweysSansonWoutersen14} and set 
\begin{equation*}
y_{i}=D_{i}\theta ^{\ast }+u_{i},\qquad D_{i}=z_{1i}\pi +v_{2i},
\end{equation*}%
with $z_{1i},v_{2i}\sim N(0,1)$ and 
\begin{equation*}
u_{i}=\rho v_{2i}+\sqrt{1-\rho ^{2}}\,\xi _{i},\quad \xi _{i}\sim \mathcal{N}%
(0,1).
\end{equation*}%
We fix $\rho =0.5$, $\theta _{0}=100$, and we let $T=800$. We consider $%
Z_{i}=(z_{1i}^{1},\dots ,z_{1i}^{4})^{\prime }$ and choose instrument
strength corresponding to the population $F$-statistic to be equal to $10$. 

\begin{figure}[tbph]
\caption{Power curves for the exact, feasible, and infeasible LR and CLR tests
with homoskedastic errors} \label{fig:homoscedastic_errors}

\centering
\par
\begin{minipage}{\textwidth}
                \centering
                \begin{subfigure}[b]{0.4\textwidth}
                    \centering
                    \caption{LR test}
                    \includegraphics[trim = 15mm 7mm 15mm 8mm, clip, width=\linewidth]{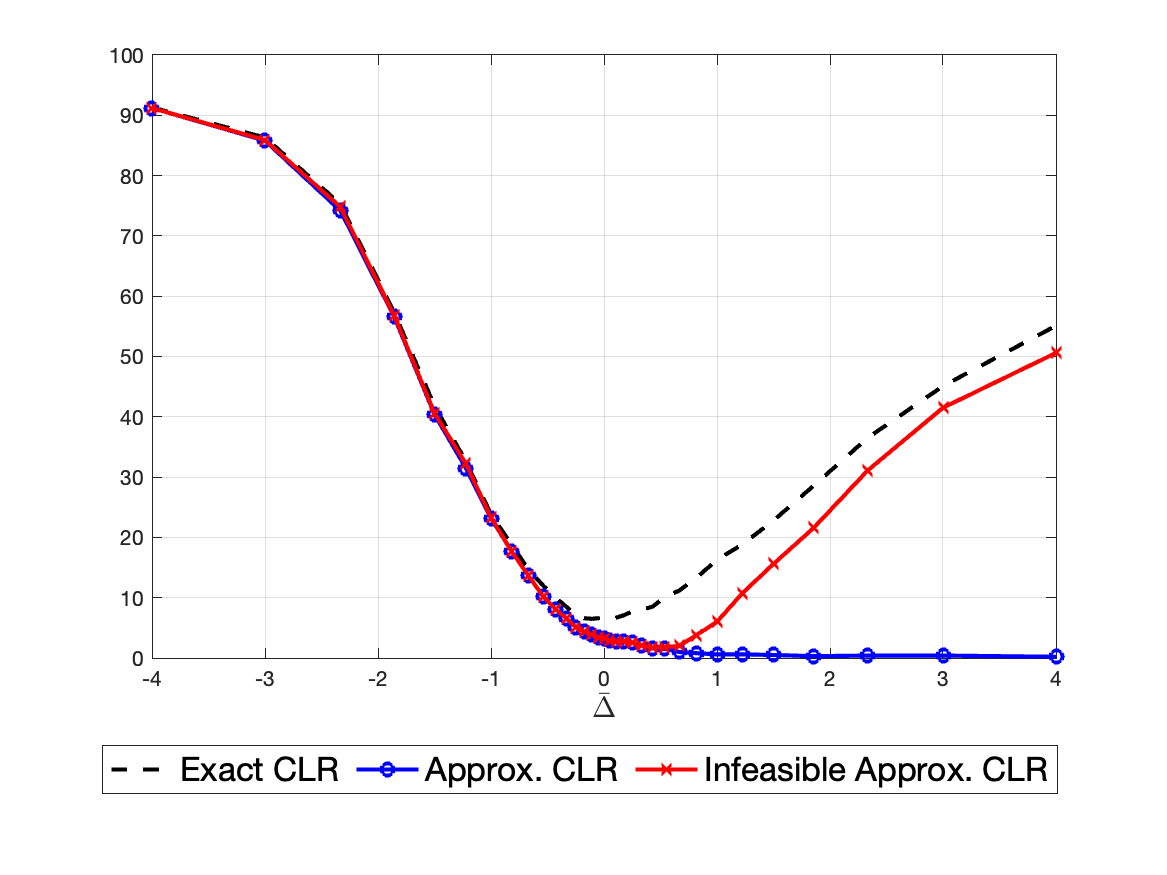}
                \end{subfigure}
                \hfill
                \begin{subfigure}[b]{0.4\textwidth}
                    \centering
                    \caption{CLR test}
                    \includegraphics[trim = 15mm 7mm 15mm 8mm, clip, width=\linewidth]{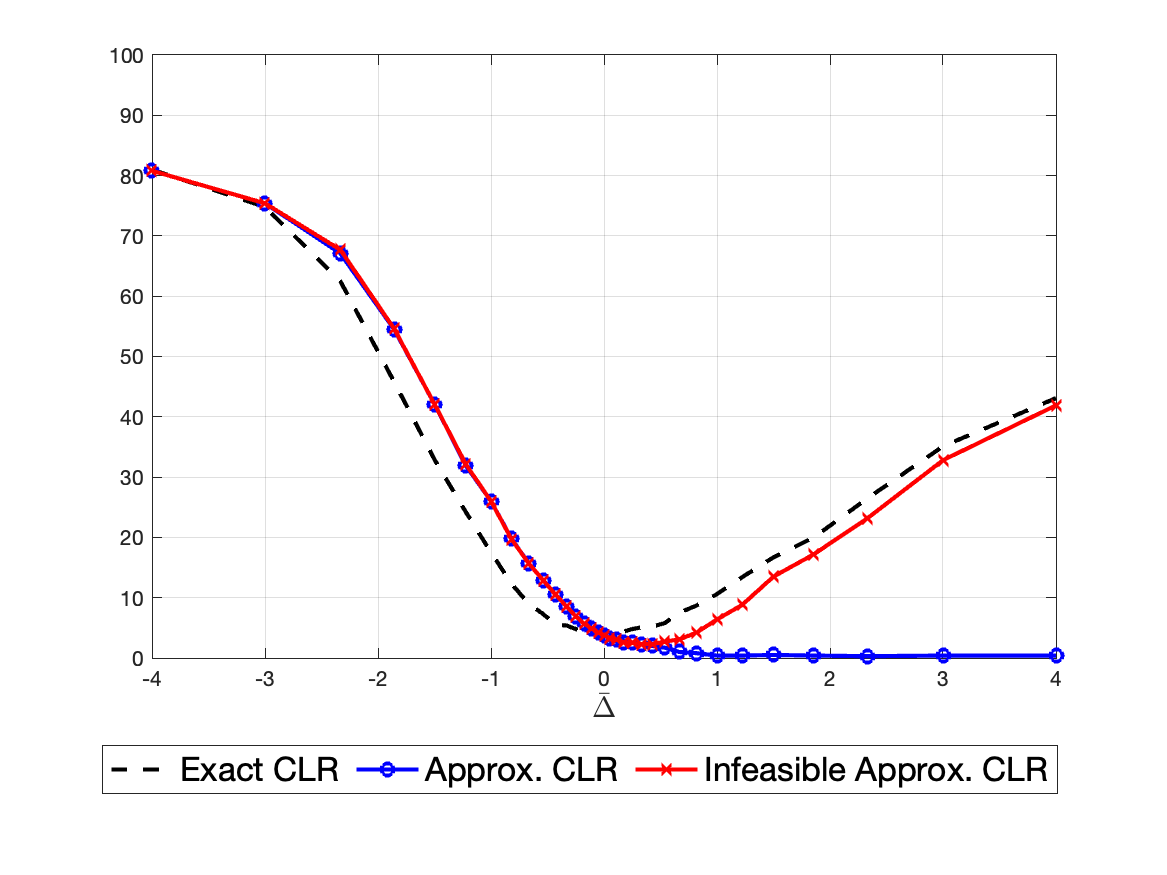}
                \end{subfigure}
            \end{minipage}
\end{figure}

For this design, we compare the power curves of the LR and CLR tests using the set of over four thousand points used in the supplement of \citet*{AndrewsMikusheva16} for the discrete approximation algorithm to the curves produced by the exact eigenvalue method.\footnote{We are grateful to Isaiah Andrews for generously sharing their codes.} The online supplement of  \citet*{AndrewsMikusheva16} provides a natural benchmark for grid choice. Their grid is publicly available, developed independently of our analysis and, as \citet*{MoreiraSharifvaghefi26} show, it performs well in their simulation designs. For the LR test, the critical value is the 95\%
quantile of the $\chi ^{2}(1)$ distribution, while for the CLR test we use
simulated critical values with 1000 replications. The resulting power curves
in Figure \ref{fig:homoscedastic_errors} highlight the differences between the discrete approximation and the eigenvalue
method. The values on the x-axis are the scaled difference between $\theta $ and $\theta_{0}$ given by
$\bar{\Delta} = (\theta - \theta_{0}) 
\left\Vert \Sigma_{0}(\theta_{0},\theta_{0})^{-1/2} \mu \right\Vert$,
where $\mu = \sqrt{n} \pi$, $\Sigma_{0}(\theta_{0},\theta_{0}) = \left[ b(\theta_{0} )^{\prime
}\otimes I_{k}\right] \Sigma_{0}\left[ b(\theta_{0})\otimes
I_{k}\right]$ and $\Sigma_{0}$ is the value of $\widetilde{\Sigma}$ under the null hypothesis.

The figures show that the feasible grid severely underestimates power for
both LR and CLR tests, often collapsing to zero near the null. The
infeasible grid improves the approximation but is not exact: for the LR test
it stays below the exact power curve, while for the CLR test it can exceed
the exact power in some regions. \citet*{AndrewsMoreiraStock06} show that the
CLR test is nearly uniformly most powerful within three classes of tests:
asymptotically efficient, two-sided invariant, and unbiased tests. It follows that
the infeasible CLR test can yield artificially inflated power. This
underscores both the practical limitations of feasible grids and the
potentially overly favorable impression created by infeasible ones.

\subsection{Design with HAC errors}

\label{sec:hac_simulation_designs}

We now turn our attention to the performance of the approximation method in
linear IV models with HAC errors. When the errors are not homoskedastic, the
non-convex CU--GMM objective function diverges from the form of a generalized
Rayleigh quotient. Until recently it was not clear how to properly solve the
optimization problem. \citet*{MoreiraNeweySharifvaghefi24} show
that the CU--GMM objective function in linear IV settings takes the form of a
ratio of polynomials with respect to the structural coefficient, $\theta $.
Hence, when there is one endogenous variable, the optimization problem can
be translated into finding the roots of a polynomial equation of order at
most $4k-1$, which can, in turn, be transformed into an eigenvalue problem.
Here, we compare the behavior of the approximated CLR test with the exact CLR
test, based on the polynomial method of \citet*{MoreiraNeweySharifvaghefi24}.

To cover a wide range of designs with HAC errors and to avoid relying on a specific setup to model heteroskedasticity and autocorrelation among the errors, we generate data from the transformed reduced-form model considered by \citet*{MoreiraMoreira19} for the linear IV regression with one endogenous variable: 
\begin{equation}
R=\mu a(\theta ^{\ast })^{\prime }+V.  \label{eq: reduced_form_model}
\end{equation}%
Here, $R=\left( Z^{\prime }Z\right) ^{-1/2}Z^{\prime }Y$, $\mu =\left(
Z^{\prime }Z\right) ^{1/2}\pi $, and $a(\theta ^{\ast })=(\theta ^{\ast
},1)^{\prime }$. The $k\times 2$ matrix of transformed reduced-form errors, $%
V$, has a normal distribution with mean zero and $2k\times 2k$ variance
matrix $\Sigma$. By Theorem \ref{thm: size_approx_clr_test}, when 
\begin{equation}
\min_{\theta \in \Theta _{P_{T}}\backslash \{\theta _{0}\}}\left\{ |\theta
-\theta _{0}|.\Vert \Sigma (\theta ,\theta )\mu \Vert \right\} ,
\label{eq:cond_size_failure}
\end{equation}%
where $\Sigma (\theta ,\theta )=\left( b(\theta )^{\prime }\otimes
I_{k}\right) \Sigma \left( b(\theta )\otimes I_{k}\right) $ is large
(diverging to infinity as $T\rightarrow \infty $), the approximated CLR test
fails to be similar (and is consequently biased). But how large does the term in (\ref{eq:cond_size_failure}) need to be in practice for size distortions to arise when using the illustrative set of points employed in the simulation study of \citet*{AndrewsMikusheva16} for the approximated CLR test?

To answer this question, we generate 500 random designs that cover a wide
range of values for the term given by (\ref{eq:cond_size_failure}). Each
design corresponds to a value of $\theta ^{\ast }=\theta _{0}$ generated
from the standard two-sided L\'{e}vy distribution, $\Sigma $ generated from
the standard Wishart distribution with $2k$ degrees of freedom, and $\mu
=100\cdot \delta $, where each element of $\delta $ is generated from the
standard two-sided L\'{e}vy distribution. These choices are meant to guarantee
quite different generated values for the reduced-form parameters. Out of the
500 simulated values of $\theta _{0}$, only two corresponding to $k=10$ and
one to $k=50$ lie outside the grid range considered by \citet*{%
AndrewsMikusheva16}. These designs are marked as orange circles. 
\begin{figure}[h]
\caption{{\protect\small Comparing the size for the exact CLR test with those
generated by the discrete approximation algorithm in the linear IV
regression with HAC errors.}}
\label{fig:size_distortion_hac_iv}\centering
\par
\centering
\begin{minipage}{\textwidth}
        \centering
        \begin{subfigure}[b]{0.45\textwidth}
            \centering
            \caption{}
            \includegraphics[trim = 0mm 0mm 0mm 0mm, clip, width=\linewidth]{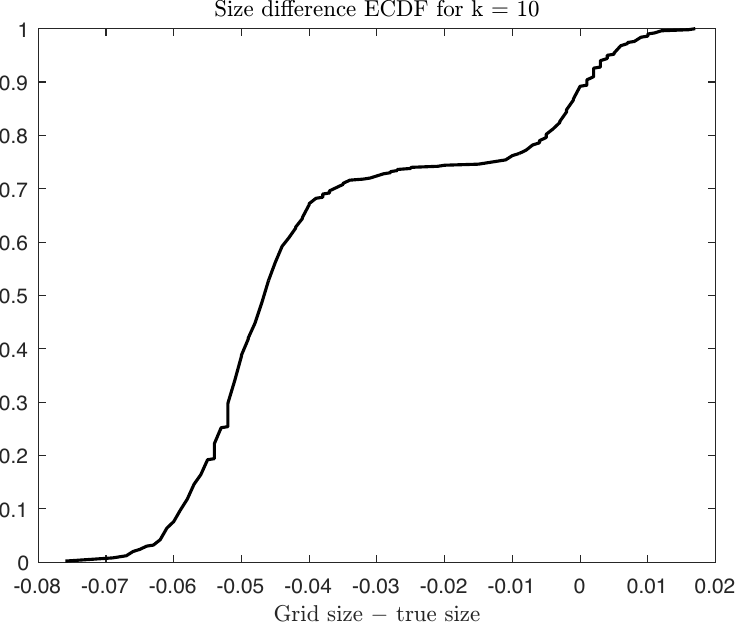}
        \end{subfigure}
        \hfill
        \begin{subfigure}[b]{0.45\textwidth}
            \centering
            \caption{}
            \includegraphics[trim = 0mm 0mm 0mm 0mm, clip, width=\linewidth]{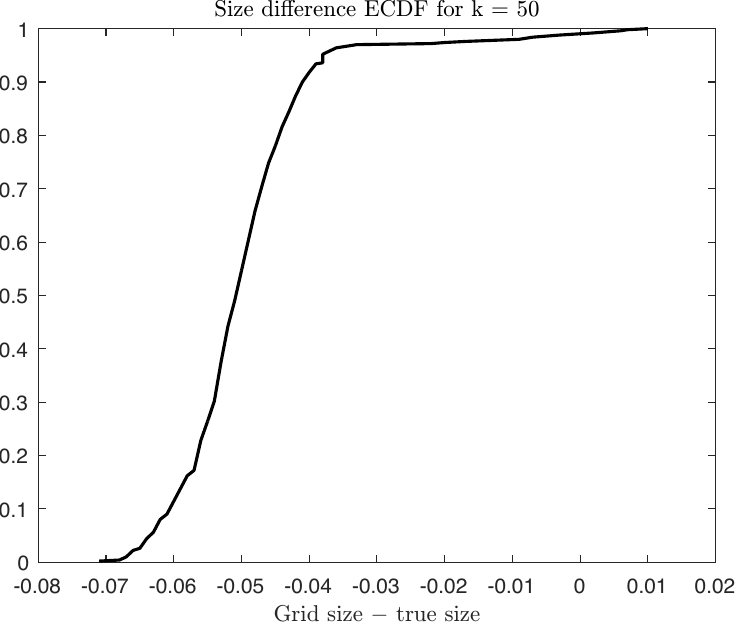}
        \end{subfigure}

    \end{minipage}
\par
\begin{minipage}{\textwidth}
        \centering
        \begin{subfigure}[b]{0.45\textwidth}
            \centering
            \caption{}
            \includegraphics[trim = 0mm 0mm 0mm 0mm, clip, width=\linewidth]{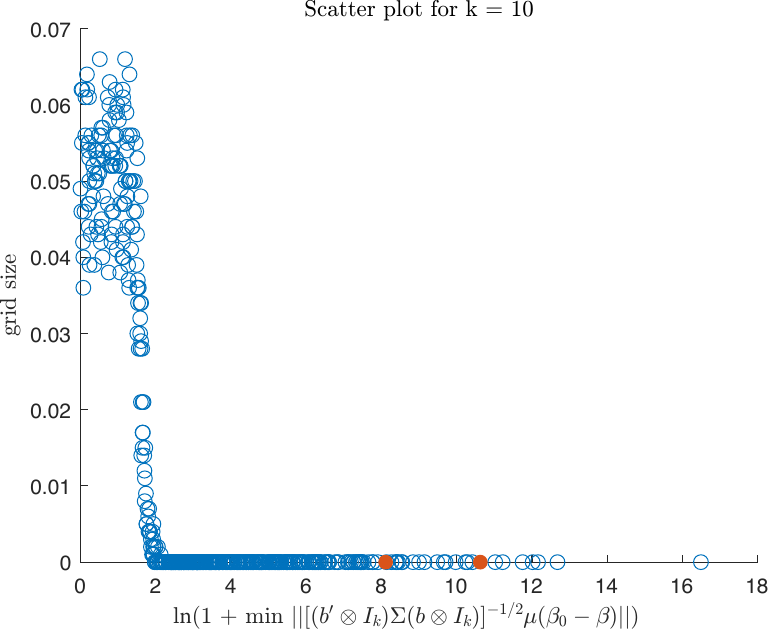}
        \end{subfigure}
        \hfill
        \begin{subfigure}[b]{0.45\textwidth}
            \centering
            \caption{}
            \includegraphics[trim = 0mm 0mm 0mm 0mm, clip, width=\linewidth]{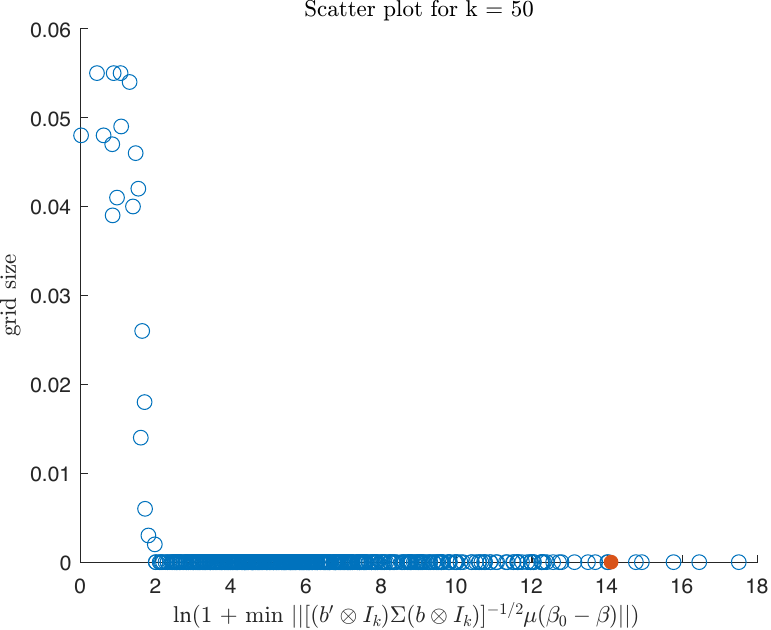}
        \end{subfigure}
    \end{minipage}
\par
\end{figure}

The plots in the first row of Figure \ref{fig:size_distortion_hac_iv} show
the empirical cumulative distribution functions (CDFs) based on these 500 designs for the difference in computed
size between the exact CLR test and the approximated CLR test when $k=10$
(panel (a)) and $k=50$ (panel (b)). The size of the implemented exact CLR test is very close to the nominal 5\%. The small deviations around 5\% are due to simulation noise arising from the finite number of replications used to approximate the null distribution. For the grid-based CLR, the probability of a Type I error varies between 0\% and about 6–7\%. When $k=10$, in more than $80$ percent
of the designs, the size of the approximated CLR test is below that of the
exact CLR test. This proportion increases to more than $90$ percent when $%
k=50 $.

The plots in the second row show that when 
\begin{equation*}
\ln (1+\min_{\theta \in \Theta \backslash \{\theta _{0}\}}\left\{ |\theta
-\theta _{0}|.\Vert \Sigma (\theta ,\theta )\mu \Vert \right\} )
\end{equation*}%
exceeds $1.5$, the approximated CLR test starts to fail to control size.
When the value exceeds $2$, the approximated CLR test has zero size, while
the exact CLR test always controls size at the target level of $5\%$.

\begin{figure}[h]
\caption{{\protect\small Comparing the power curves for the exact CLR test
with those generated by the discrete approximation algorithm in the linear
IV regression with HAC errors.}}
\label{fig:power_distortion_hac_iv}\centering \centering
\par
\begin{minipage}{\textwidth}
        \centering
        \begin{subfigure}[b]{0.45\textwidth}
            \centering
            \caption{k=10}
            \includegraphics[trim = 0mm 0mm 0mm 0mm, clip, width=\linewidth]{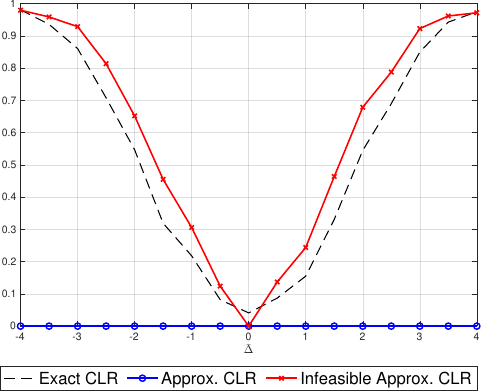}
        \end{subfigure}
        \hfill
        \begin{subfigure}[b]{0.45\textwidth}
            \centering
            \caption{k=50}
            \includegraphics[trim = 0mm 0mm 0mm 0mm, clip, width=\linewidth]{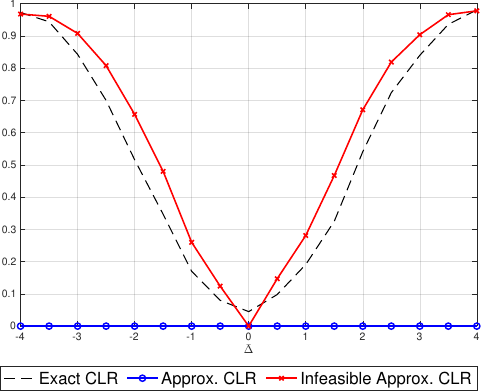}
        \end{subfigure}

    \end{minipage}
\par
\begin{minipage}{\textwidth}
        \centering
        \begin{subfigure}[b]{0.45\textwidth}
            \centering
            \caption{k=10}
            \includegraphics[trim = 0mm 0mm 0mm 0mm, clip, width=\linewidth]{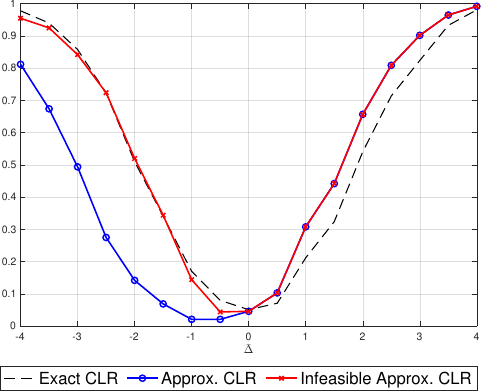}
        \end{subfigure}
        \hfill
        \begin{subfigure}[b]{0.45\textwidth}
            \centering
            \caption{k=50}
            \includegraphics[trim = 0mm 0mm 0mm 0mm, clip, width=\linewidth]{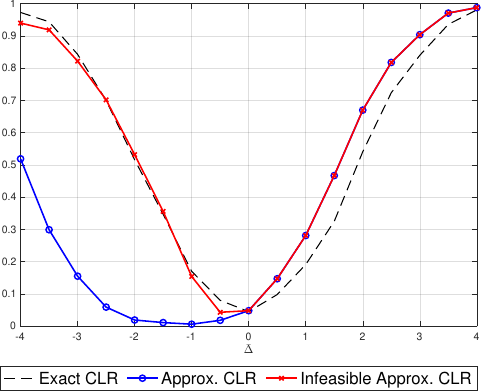}
        \end{subfigure}
    \end{minipage}
\end{figure}

We now illustrate the potential power problems even when the null rejection
probability is close to $5\%$. In the first row of Figure~\ref%
{fig:power_distortion_hac_iv} we show the power curves for both the feasible
and infeasible approximated CLR methods, considering one of the current 500
random designs that has a large value above $4.5$ for 
\begin{equation}
\ln \left( 1+\min_{\theta \in \Theta \backslash \{\theta _{0}\}}\left\{
|\theta -\theta _{0}|.\Vert \Sigma (\theta ,\theta )\mu \Vert \right\}
\right) 
\end{equation}%
when $k=10$ (left panel) and $k=50$ (right panel), where the size problem
has already been observed. In the second row of Figure~\ref%
{fig:power_distortion_hac_iv}, we further consider an alternative version of
these designs in which the parameter $\mu $ is rescaled by a constant such
that 
\begin{equation}
\ln \left( 1+\min_{\theta \in \Theta \backslash \{\theta _{0}\}}\left\{
|\theta -\theta _{0}|.\Vert \Sigma (\theta ,\theta )\mu \Vert \right\}
\right) =1.25,
\end{equation}%
and, given the plots provided in Figure~\ref{fig:size_distortion_hac_iv}, we
expect no size distortion.

In the first row, we observe that the power is equal to zero for the
feasible version of the approximated CLR test, even when distinguishing the
null from the alternative is easy (the exact CLR test achieves power equal to
one). We also observe that including the true value of $\theta$, denoted $%
\theta^{\ast}$, in the search set for the approximated CLR test, which
renders the test infeasible, dramatically changes its behavior and can
spuriously produce power exceeding that of the exact CLR test. The plots in
the second row show that, even in regions where the points in
the search set are sufficiently dense to ensure no size distortion, both the
feasible and infeasible CLR tests can still exhibit distorted power. This distortion is more pronounced for the feasible version.

\subsection{Simulation study based on IES application}

\label{sec:ies_simulation_designs}

\Citet*{Yogo04} investigates the inference problem for the intertemporal
elasticity of substitution (IES) coefficient in linear instrumental
variables (IV) regressions that may suffer from weak instruments. The study
analyzes four alternative specifications: real annual consumption growth as
the dependent variable with the real annual interest rate as the endogenous
regressor; the reverse specification, where the real interest rate is the
dependent variable and consumption growth is endogenous; and two additional
IV regressions that replace the real interest rate with the real aggregate
stock return. For each specification, the IES coefficient and its
corresponding confidence interval are estimated for 11 developed countries.

\Citet*{AndrewsMikusheva16} conduct a simulation study based on the dataset of \Citet*{Yogo04}. In their simulations, they treat the consistent HAC estimator of the variance matrix of the reduced-form errors in Equation~(\ref{eq: reduced_form_model}) as the true value of~$\Sigma$. They also take the empirical estimate of~$\mu$, $\widehat{\mu}$, as its true value, implementing the CLR test through a discrete approximation method.

However, when the instruments are weak,~$\mu$ is not consistently estimable, and different choices of~$\mu$ can lead to substantially different size and power calculations. It is of course possible to select a particular value of~$\mu$ for a given~$\Sigma$ that drives both size and power close to zero. A more interesting exercise is to consider the case of a researcher who does not know~$\mu$ and must estimate it. With high probability, the true value will lie within the confidence region around~$\widehat{\mu}$. The natural question, then, is: among the values in this confidence region, can we find a~$\mu$ that generates even more distortions?

The simulations show that power distortions can indeed arise, even when the grid is dense and rejection probabilities appear close to nominal under a particular calibration. To illustrate this sensitivity, we take the estimate~$\widehat{\mu}$ from the \citet*{Yogo04} dataset and numerically search over its confidence interval to identify a value~$\widetilde{\mu}$ that maximizes Expression~(\ref{eq:cond_size_failure}). We then treat~$\widetilde{\mu}$ as the true parameter in our simulations.

We focus on Design 1 for France following \citet*{MoreiraRidderSharifvaghefi26}
and analyze how power distortions vary across designs. Figure \ref%
{fig:power_distortion_ies_hac_iv} presents the power curves obtained using
both IV coefficients, $\widehat{\mu }$ and $\widetilde{\mu }$. No
meaningful size distortions appear for these specific designs for the grid chosen
by \citet*{AndrewsMikusheva16}. However, power distortions do arise, even
though the grid was chosen to accommodate this specific design. The extent
of the power discrepancies between the exact CLR and both the feasible and
infeasible CLR tests varies with the simulated $\mu $. Both the feasible and
infeasible approximated CLR tests yield misleading results relative to the
exact CLR test. Their power functions sometimes fall below and at other times
exceed the exact power. For certain alternatives, the discrepancy between the
exact and approximated CLR tests exceeds 10 percentage points. 
\begin{figure}[tbph]
\caption{{\protect\small Comparing the power curves for the exact CLR test
with those generated by the discrete approximation algorithm in designs
based on IES application.}}
\label{fig:power_distortion_ies_hac_iv}\centering
\par
\begin{minipage}{\textwidth}
        \centering
        \begin{subfigure}[b]{0.45\textwidth}
            \centering
            \caption{Using $\mu = \widehat{\mu}$ for France}
            \includegraphics[trim = 15mm 7mm 15mm 8mm, clip, width=\linewidth]{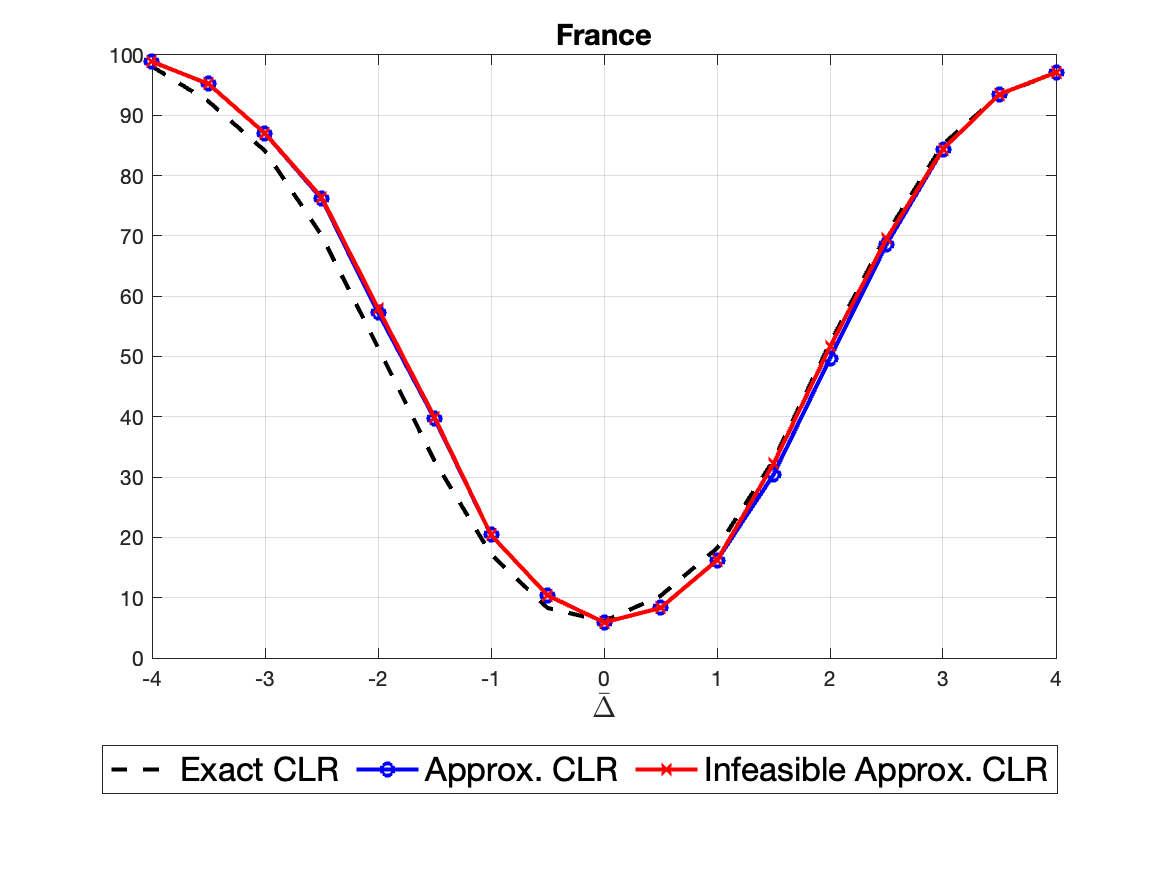}
        \end{subfigure}
        \hfill
        \begin{subfigure}[b]{0.45\textwidth}
            \centering
            \caption{Using $\mu = \widetilde{\mu}$ for France}
            \includegraphics[trim = 15mm 7mm 15mm 8mm, clip, width=\linewidth]{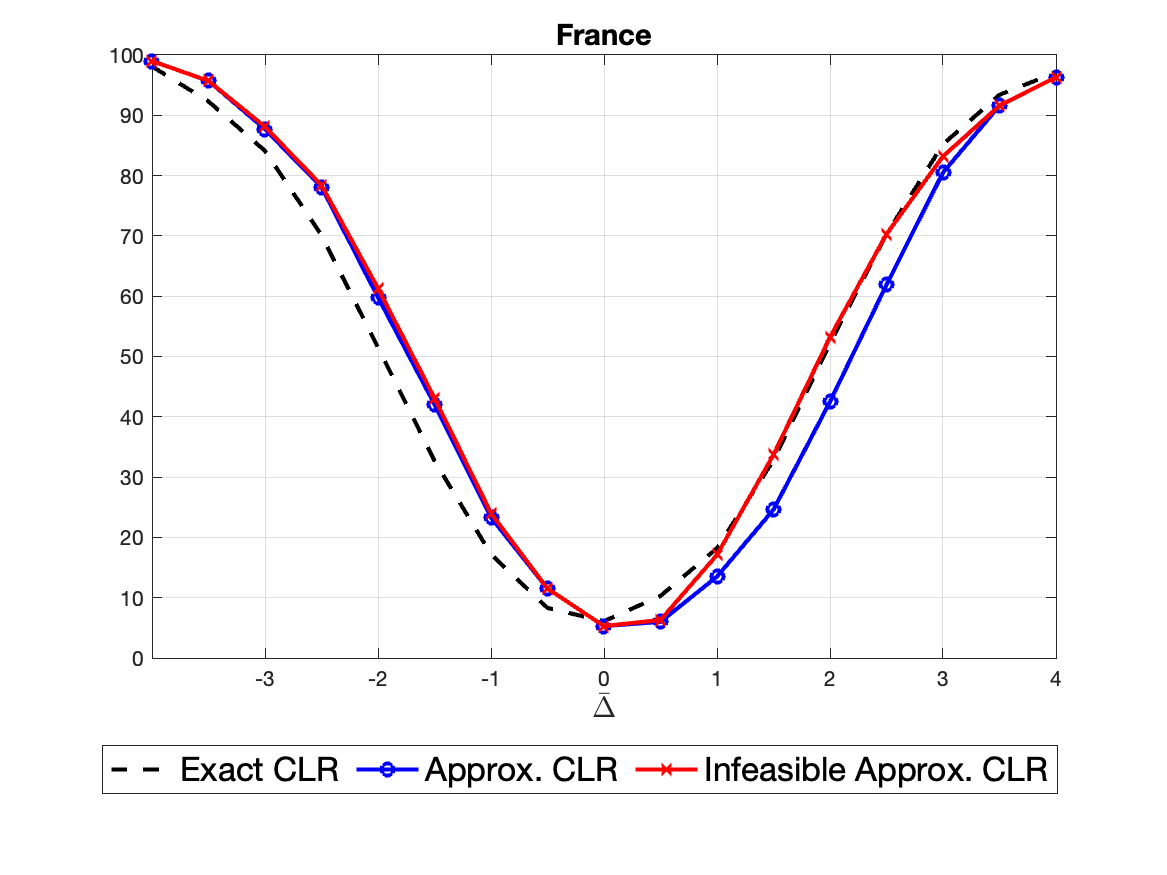}
        \end{subfigure}
\end{minipage}
\end{figure}

Even when size distortions are effectively
eliminated, as in the carefully chosen fine grid search set of \citet*{AndrewsMikusheva16},
the approximate implementations of the CLR test may still display sizable
power distortions. Both the feasible and infeasible approximations can yield
a misleading representation of the exact power function, at times overstating
and at other times understating it. These discrepancies indicate that the
reliability of approximate CLR procedures depends not only on numerical
precision but also on the choice of nuisance parameter values. Consequently,
accurate inference under weak identification requires attention to both the
control size and power properties across different designs.

\section{Conclusion}

The CLR test remains a valuable tool for inference under weak identification,
with appealing theoretical properties in both linear and nonlinear settings.
Its implementation, however, is complicated by the need to minimize a
non-convex CU--GMM objective function. Our paper emphasizes that practical implementation requires careful attention to numerical methods. 
Discrete approximation strategies that simplify optimization are not merely a computational detail. They can distort the test statistic, the conditional critical value function, and consequently the size and power of the CLR test. While grid-based procedures may perform well in particular designs, they do not provide uniform guarantees that the resulting approximated CLR test preserves the theoretical properties of the exact CLR test, even when the number of grid points increases at a polynomial rate in the sample size. Recent progress in the linear IV case by \Citet*{MoreiraNeweySharifvaghefi24} shows that polynomial methods can deliver exact global minimization, thereby bridging the gap between theory and computation.

In the nonlinear setting, implementation can be based on piecewise polynomial or polynomial-ratio approximations. Guaranteeing that the theoretical properties of the CLR test remain valid in practice, however, requires a new theoretical framework. Current joint work extends these ideas to more general models, with the goal of delivering implementations that are both theoretically robust and computationally reliable.


\bibliographystyle{chicago}
\bibliography{References}


\end{document}